*Review Article*

# Retrograde versus Prograde Models of Accreting Black Holes

## David Garofalo

*Department of Physics, Columbia University, New York, NY 10027, USA*

Correspondence should be addressed to David Garofalo; david.a.garofalo@gmail.com

There is a general consensus that magnetic fields, accretion disks, and rotating black holes are instrumental in the generation of the most powerful sources of energy in the known universe. Nonetheless, because magnetized accretion onto rotating black holes involves both the complications of nonlinear magnetohydrodynamics that currently cannot fully be treated numerically, and uncertainties about the origin of magnetic fields that at present are part of the input, the space of possible solutions remains less constrained. Consequently, the literature still bears witness to the proliferation of rather different black hole engine models. But the accumulated wealth of observational data is now sufficient to meaningfully distinguish between them. It is in this light that this critical paper compares the recent retrograde framework with standard "spin paradigm" prograde models.

## 1. Introduction

When Roy Kerr presented his solution at the Texas Symposium almost five decades ago, the astronomical community, ironically, was too busy with the recent discovery of quasars to pay attention. But the importance of black holes is now grounded in observations pointing to a breadth and depth of a black hole impact that is likely still underestimated. In galaxies, black holes appear to participate in triggering and in quenching star formation, in heating and expanding gas, and in altering the mode of accretion [1–7]. They are connected to bulges and stellar dispersions at the spatial extremes of galaxies [8, 9], and in many cases the impact of black holes appears on cluster environments as well [1, 10–13]. But black hole influence is also observed on smaller scales, with stellar- mass black holes producing a rich panoply of observational signatures [14].

Black holes are both spatially and gravitationally irrelevant to galaxies as a whole, so the influence they exert is thought to occur during active phases when large amounts of energy spew from their centers in both kinetic form and radiation spanning the entire electromagnetic spectrum. Because the black hole scaling relations are ubiquitous, perhaps all galaxies experience an active phase during which black holes reveal their presence to the larger galaxy. And this active phase would involve accretion onto supermassive black holes [15, 16], the formation of powerful winds [17], and in some cases jet formation [18, 19], but the details of how jets are related to accretion remain elusive.

The earliest analytic models involve accretion in either thin-disk or advection-dominated form [15, 16, 20, 21],

and/or spin energy extraction from black holes ([18]; henceforth BZ). Because of the nature of the relativistic gravitational potential, the thin-disk accretion model [16] is characterized by the existence of stable circular orbits with an inner boundary inwards of which gas plunges rapidly onto the black hole [22]. This innermost stable circular orbit (ISCO) depends on the spin of the black hole and the relative orientation of the black hole angular momentum relative to the angular momentum of the accretion flow. When the two angular momenta are aligned and the magnitude of the black hole angular momentum (or spin) is large, the accretion disk lives closer to the black hole which translates into a greater accretion efficiency (up to $0.42\ \dot{M}c^2$, with $\dot{M}$ as the accretion rate onto the black hole). This is due to the presence of accretion material close in to the black hole where it can tap into the strong gravitational potential and reprocess that energy further into the disk. As the black hole spin value drops toward zero, the accretion efficiency drops as well, reaching $0.06\ \dot{M}c^2$ at zero spin, due to an ISCO located further away from the black hole, so less energy is reprocessed into the disk. In the retrograde regime, where the black hole's angular momentum is opposite that of the accretion flow (i.e., they rotate in opposite directions), the accretion efficiency is even lower than that of the zero spin. Unsurprisingly, this is due to an ISCO that moves even further outwards.

The efficiency of the jet production in the BZ model depends on the spin value of the black hole and the strength of the magnetic field surrounding the black hole via

$$L \propto B^2 a^2, \qquad (1)$$

where $B$ is the magnetic field strength on the black hole and $a$ is the spin parameter which is a dimensionless number (varying in magnitude from 0 to 1) characterizing the angular momentum of the black hole. Therefore, the disk and jet efficiency scale directly with prograde black hole spin (i.e., they both increase or decrease together). This framework is referred to in the literature as the "spin paradigm" [23–26] and, as we shall see when confronted with observation, is fraught with a host of difficulties, most of them emerging in the last few years.

Due to jet efficiency issues in the spin paradigm, hybrid spin paradigm models have been developed that in addition to jets from black holes include the contribution of jet outflow from the disk ([19]; henceforth BP) and the ability of the ergosphere to enhance the strength of the black hole-driven jet [27–29]. Hybrid spin paradigms explore the ramifications of high/low black hole spin surrounded by thin- disk or advection-dominated accretion in an attempt to enhance the jet efficiency. These ideas have been in part a response to arguments against powerful jets in black hole systems due to diffusive effects [30–32], suggesting that the black hole-threading magnetic field would be too weak for the BZ mechanism to explain the observations. This has also been a concern of general relativistic magnetohydrodynamic simulations (GRMHD) which we will discuss in some detail [33–36].

While the most radical departure from the spin paradigm, referred to as the "gap paradigm" [37], is also constructed from the basic building blocks of the other models (including BZ, BP, and thin/thick disks), it also, and crucially, hinges on the ability of retrograde accretion to produce the most powerful jets. This framework, therefore, argues that in addition to high/low black hole spin and thin-disk/advection dominated accretion, retrograde/prograde directions between disk and black hole also matter.

This paper is not an introduction to black hole engines (for an exhaustive review and detailed treatment see [38]). Its purpose is to highlight the differences between standard prograde models of accreting black holes and the recent retrograde framework. And because the track record on the discussion of compatibility between theory and observation has been poor, that is the focus of this paper.

## 2. On Supermassive Black Hole Formation

The underlying assumption of the black hole paradigm for AGN that supermassive black holes are produced in galactic centers is not a completed story because the mechanism for producing $10^9$ solar mass black holes at early times remains unknown. Specifically, if the high redshift FRII radio quasars are modeled as the result of cold gas accretion onto black holes [39] with nonzero spin, constraints are imposed on the black hole formation scenarios. While models of spiral galaxies suggest that central black hole accretion is produced by secular processes [40–43] onto smaller-sized black holes (i.e., $10^5$–$10^7$ solar masses) that originate from primordial seeding mechanisms that are still debated, the supermassive black holes in the centers of elliptical galaxies are presumed to be the product of galaxy mergers, where two already massive black holes come together. However, merging black holes suffer the so-called "last parsec problem" [44]. Because the two black holes are unlikely to fall straight into each other but find themselves orbiting one another, a successful merger comes about only if the binary angular momentum can be extracted. While the conventional wisdom has been that the excess angular momentum ends up in stars and gases that are close to the center of the newly formed galaxy, a problem emerges. Stars and gases may well acquire the angular momentum, but they are then pushed outwards or away from the black holes, leaving the two black holes isolated, without a complete repository for shedding the remaining angular momentum required to merge the black holes. As a result, the merger process stalls at a characteristic distance of about 1 pc, well short of the 0.01 pc necessary for gravitational waves to complete the job [44]. While signatures of binary black holes are observed [45], it appears to be only a small fraction (about 10%) in the local universe but very small (about 0.3%) for quasars at all redshifts [46]. For the largest black holes, the merger process generally appears to operate to completion. How does this happen?

While a number of possible mechanisms have been proposed, recent numerical simulations of retrograde accretion onto binary black holes show that the efficiency of this process in extracting both the energy and angular momentum of the binary is greater than that of a prograde disk [47]. Here, the angular momentum of the accretion flow is antiparallel to that of the black hole binary, that is, the accretion flow is rotating in the opposite direction compared to the direction in which black holes rotate about one another. Because angular momentum acquired by the retrograde flow causes the accretion flow to move further inward toward the binary, unlike in the prograde case, a repository for the angular momentum of the binary continues to be present throughout the merging process, and the black holes successfully shed their binary angular momentum as they approach the sub-parsec regime.

Interestingly, retrograde accretion onto a massive black hole is the starting assumption of the gap paradigm [37]. However, simulations suggest that the postmerger black hole will tend to have low retrograde spin [47], which imposes hitherto unclear constraints on the gap paradigm. Given that the strengths of magnetic fields threading black holes are unknown, the extent to which highest spinning retrograde black holes (as opposed to intermediate retrograde spin values) are needed is weakly constrained. Once magnetic field strengths in the very central regions are better estimated observationally, the model predictions can be squared with the observed redshift distribution of radio loud quasars. In more detail, once the simulations specify the fraction of postmerger systems producing rapid retrograde spin, using as input both the merger function versus redshift and magnetic field values in the inner accretion flow, we can determine the number of retrograde systems predicted by the model that will satisfy the observed requirements of the powerful FRII radio quasars, allowing the model to be tested in this respect. Spin and hybrid spin paradigm models are built on the assumption that FRII quasars have high prograde black hole spin. This difference, as we shall see in the next section, is crucial.

## 3. On Flat Spectrum Radio Quasars

If spin and hybrid spin paradigm models postulate rapid spinning black holes surrounded by a thin-disk accretion generated in postmerger ellipticals with powerful jets in a prograde configuration, the ISCO is closer to the horizon. Interestingly, the evidence in Flat-Spectrum Radio Quasars (FSRQs) points to ISCO values that are not close to the black hole but compatible with retrograde accretion [48–50]. This issue is addressed in spin paradigm and hybrid spin paradigm models via the added assumption that FSRQs are rapidly accreting systems, which may produce radiatively inefficient, geometrically thick disks [20, 31, 51, 52]. The purpose of this additional assumption is twofold. First, there is a need to explain the absence of strong X-ray reprocessing features from the inner regions, which would exist under the assumption of a radiatively efficient, geometrically thin disk plus hot corona models [53] within the context of high prograde spin. Second, numerical simulations have suggested that strong jets are produced only in geometrically thick accretion systems, but there is tension here. Radiatively inefficient accretion will tend to wash out or be incompatible with prominent signatures of thermal accretion such as broad optical lines and big blue continuum bumps, which means that as you consider thicker disk geometry, you tend to weaken the thermal disk signatures, but the absence of evidence for ISCOs close to the black hole come from the possibility of modeling the system via radiatively efficient thin disk accretion. In fact, the thermal spectra in such objects produce big blue bumps which are strong signatures of radiatively efficient thin-disk accretion with maximum disk temperatures lower than expected [54]. That tension is avoided in the gap paradigm since the most powerful jets in that framework are produced in thin-disk systems with largest gap regions where the ISCO is further out from the black hole, and the less reprocessed energy through the disk is compatible with a lower peak frequency for the big blue bump compared to radio quiet quasars [55]. In high prograde spin models, the need for both advection-dominated accretion as well as thin disks to model the observations suggests an additional way out, namely disk truncation. Here, the radiatively efficient nature of thin disks abruptly ceases to operate at some location in the disk, giving rise to radiatively inefficient thick-disk geometry inwards of that location [56]. However, this picture produces tension with the assumption of scale invariance (discussed in Section 4) because X-ray binaries in either soft states or transitory burst states appear not to be the small-scale versions of FRII quasars. Finally, we should point out that recent work suggests the absence of jet power dependence on disk thickness [57], which removes the necessity to associate powerful jets with radiatively inefficient accretion.

### 4. On FRII Radio Quasars and Ballistic Microquasar Jets

Although from a theoretical perspective the detailed physics such as accretion disk temperature and density and outflow power do depend on black hole mass, the mechanism for producing that outflow does not. In other words, the nature of the mechanism producing the outflow is the same regardless of the size of the accreting black hole. This so-called "scale invariance" produces additional constraints on the modeling of the most powerful sources with jets—FRII radio quasars— and their time evolution. To see this, we must compare with their small-scale counterparts.

Stellar mass black hole accretion systems—X-ray binaries [58]—undergo transitions between states with jets (with relativistic gamma factors whose value is less than about 2), dominated by a hard X-ray spectrum and modeled as radiatively inefficient advection dominated accretion, and states without jets but with softer X-rays that can be well-fit by optically thick, geometrically thin disk models. And the evolutionary cycle involves a progression from hard state jets to jet-less soft states and back, with a transitory but more powerful and collimated jet outflow (with a relativistic gamma factor whose value is greater than 2) produced in the hard to soft transition [59]. While scale invariance does not require that AGN jets follow this specific time evolution, it does imply large-scale analogs among AGN. In particular, the brightest hard state jets observed in X-ray binaries may be the small-scale counterpart to the FRI radio galaxies; the lower-luminosity hard state X-ray binaries may be the small-scale version of the low-luminosity AGN, while the jet-less soft state in X-ray binaries may be the small-scale counterpart to quasars and radio quiet AGN [60–62].

Explanations of the nature of the powerful transitory jets in X-ray binaries have been explored in terms of an unstable transition between hard state and soft state [20, 63]. And the interpretation of these states has emerged from the suggestion that lifetimes of FRII radio quasars are as short compared to lifetimes of FRI radio galaxies to the same degree, as transitory jets are to hard state jets in X-ray binaries. While the underlying physical model for the generation of these transitory jets is not in itself problematic, its straightforward time evolution scale-free extension to AGN violates the observations. The cyclical nature of X-ray binary state transitions, in fact, would imply that on average there would be no straightforward redshift dependence in the density of FRI versus FRII objects. Observationally, however, FRII quasar density peaks at higher redshifts (about $z = 2$), while the density of FRI radio galaxies and black hole masses increases toward lower redshifts [64, 65].

Alternatively, one could choose to explore a scale-invariant violating framework. Although this would circumvent the constraints discussed above, no successful scale-free violating models currently exist. In fact, recently discovered powerful jets in spiral galaxies (gamma ray loud Narrow Line Seyfert 1s) with lower mass black holes [66] further strain the tension between the scale-free spin paradigm and observations by forcing us to revisit models in which jets avoid spirals. It is worth pausing at this point to emphasize the degree to which the spin paradigm is grounded in the notion that spiral galaxies have low black hole spin and cannot thus produce powerful jets [26, 67, 68]. The scale-invariant gap paradigm predicts that jets in spiral galaxies exist in lower prograde spin black hole systems in radiatively efficient accretion states [37]. Hence, it would be model-constraining to apply the broad iron line fluorescence method to such systems [69].

## 5. On the Efficiency Requirements in Powerful Radio Galaxies

With the exception of the early and isolated pioneering work of Wilson [70], GRMHD simulations have been exploring black hole accretion for the past decade. While early numerical work supported the basic analytic BZ framework by showing that energy extraction from black holes is possible [33, 35, 71, 72], GRMHD simulations have also recently begun addressing the problems discussed above concerning jet efficiency. The first GRMHD simulations were disappointing in this respect because they produced jet powers considerably less than $\dot{M} c^2$ (by jet efficiency we mean the ratio of jet power to accretion power). Recent observations, in fact, showed that in many powerful radio galaxies in hot cluster environments, the jet power is at least an order of magnitude greater than its accretion power [73–75], in some cases about 25 times the accretion power [76]. In an attempt to address this problem, recent numerical work has focused on initial conditions for the magnetic fields that produce a greater advection of magnetic flux on the black hole [77, 78]. These "flooded" magnetospheres have enhanced efficiencies that increase with prograde values of spin up to about 3 $\dot{M} c^2$ at the highest spin values (Tchekhovskoy and McKinney 2012). While this is an improvement, we are still a factor of 10 or more shy of explaining the powerful radio galaxies (e.g., the galaxy cluster RBS 797—[79]). In addition, more mild efficiency requirements in systems such as M87 also appear difficult to explain in GRMHD (even in the context of the ad hoc assumption of "flux flooding") as VLBI imaging of the central region of M87 indicates the possibility of black hole spin values around 0.6 [80]. In fact, jet efficiency in GRMHD is a steep function of black hole spin and drops down to about 30% for intermediate values of prograde spin (Tchekhovskoy and McKinney 2012), making it difficult to explain the measured jet efficiency in M87 around 5% at the Bondi radius and therefore possibly making it orders of magnitude larger near the black hole due to outflows [81]. The gap paradigm, on the other hand, produces less stringent accretion requirements for intermediate prograde spin values [37, 82]. It is unfortunate that Doeleman et al. [80] give the false impression that the possible prograde nature of the black hole in M87 constrains current models. In fact, none of their cited references suggest that M87's supermassive black hole should be retrograde. On the other hand, detailed modeling of the behavior of low angular momentum accretion onto M87 suggests caution on black hole spin inference in this source (Das et al. 2012). More generally, the proximity of M87's jet provides a wonderful opportunity for studying the shock-producing interaction of the jet with its environment in a classic FRI radio galaxy in detail [83, 84].

## 6. On Scale Invariance and FRI Radio Galaxies

FRI radio galaxy jets are generally weaker and less collimated on kpc scales than their powerful FRII counterparts with an observed distribution captured by the so-called Owen-Ledlow diagram. We mentioned that FRI radio galaxies are thought to satisfy scale invariance as the large-scale analogs of bright hard state X-ray binaries. When this assumption is coupled with recent observations, constraints are produced. Among these are the observed absence of any clear black hole spin dependence in X-ray binaries in their hard states [85] and the claim of a spin dependence in the transient ballistic jet (Narayan and McClintock 2011).

It is worth emphasizing that the observed lack of difference in jet power between systems with different black hole spin in hard state X-ray binaries [85] does not constitute evidence that black hole spin energy extraction is not occurring—although it is compatible with that notion. It is evident that whichever process influences jet power, it produces a black hole spin dependence that is roughly flat. The possibility of a flat spin dependence is problematic for GRMHD which has consistently produced steep spin dependencies of jet power of the form $a^2$–$a^6$ [86]. While it is clear that spin cannot be the only factor in determining the presence of a jet [87], the details of the jet-disk connection remain unresolved.

What about the apparent spin dependence observed during the transition state [88]? What mechanism operates to reveal the value of black hole spin as the accretion state transitions from advection dominated to radiatively efficient and thin? Narayan & McClintock suggest a previously proposed explanation, which is that as the system moves toward the soft state and the inner edge of the disk moves inward toward the ISCO shocks may occur leading to the formation of pc scale jets [59], which would be the small-scale counterpart of the quasar jets. However, this notion breaks scale invariance because—as discussed above—it implies an absence of FRII versus FRI density dependence on redshift, which appears directly incompatible with the redshift distribution of the AGN sources.

The gap paradigm, on the other hand, is explicitly scale invariant in this respect because the nature of the radio-loud to radio-quiet transition hinges on the observational results of Neilsen & Lee [89] and is further strengthened by Ponti et al. [90], which come from X-ray binaries. And the picture goes like this: as the disk transits toward a soft state, and the inner edge moves toward the ISCO, the radiative wind component of the disk increases which serves to collimate the black hole or BZ jet. As Neilsen and Lee point out, the effect is transitory since the disk wind eventually quenches the inner jet, but these considerations have implications for a spin dependence in the jet-quenching ability. This is because the wind outflow power is proportional to the energy generated at that location in the disk, which depends on the black hole spin (the larger the prograde spin value, the greater the wind efficiency). One must conclude, therefore, that the jet-quenching ability of the disk wind will be greater for larger prograde spin values. The consequence of adopting this scale-free framework, therefore, is the existence of black hole X-ray binaries at lower prograde spin whose jets are never fully quenched. The prediction in the gap paradigm is that some X-ray binaries will have jets that are more quenched in high states (corresponding to higher values of prograde spins), and others will have jets that are less fully quenched in high states (with lower values of prograde spins). The implication of this for AGN is the evolution of systems, that if persistently radiatively efficient in their accretion state, they evolve into radio-quiet quasars or radio-quiet AGN from FRII quasar states [37]. In addition, this picture suggests an explanation for the small fraction of observed FRI quasar objects in that they occupy a small range on the black hole spin spectrum (above zero but in the lower range of the prograde values).

Therefore, while the analysis of Narayan and McClintock suggesting black hole spin evidence in X-ray binaries is compatible with the gap paradigm, I would argue that the conditions producing ballistic jets in X-ray binaries are not relevant to those in high redshift AGN jets, so they cannot be the small-scale analogs of the FRII quasar jets. In spin and hybrid spin paradigm models, no problem-free, scale-invariant analog exists for the ballistic jets. In the gap paradigm, instead, FRII quasar jets are modeled as retrograde systems, and because time evolution (discussed later) does not produce thin radiatively efficient accretion following ADAF states, the large-scale analog of the ballistic jet is not observed in nature.

### 7. On the Radio-Loud/Radio-Quiet Dichotomy

In addition to the jet-disk connection, another major unresolved issue in astrophysics is the nature of the radio-loud/radio-quiet dichotomy. Observationally, we find that only 15–20% of active galaxies are radio-loud. A scale-free extension of the observations of X-ray binary state transitions does not explain the quantitative nature of the dichotomy. But if we consider retrograde accretion as a model for radio-loud quasars, we find a space for addressing the issue. Regardless of what assumption we begin with in terms of the fraction of postmerger objects that are retrograde compared with the number of prograde ones, the time evolution will take some of the radio-loud objects and naturally evolve them into radio-quiet ones. Time evolution, in fact, will spin the retrograde black hole down to zero spin in less than $10^7$ years at the Eddington rate and after a few more $10^7$ years, the black hole spin will be intermediate prograde. For the fraction of systems that persist in their radiatively efficient mode of accretion, the gap paradigm prescribes that they will turn into radio-quiet objects. However, a fraction of these originally retrograde systems will not remain radiatively efficient in their accretion states but become ADAFs. These become FRI radio galaxies so it is only a fraction of a fraction of the original radio-loud retrograde objects that become radio-quiet quasars/AGN. In short, there is a natural mechanism in the paradigm for taking originally radio loud quasars and turning a fraction of them into radio-quiet quasars/AGN. In addition, the less massive black holes in retrograde configurations may flip to prograde configurations (discussed later), thereby taking originally radio-loud systems and turning them into radio-quiet ones. This means that although the quantitative fraction of radio-loud to radio-quiet objects depends on the specific initial conditions, a qualitative explanation for having more radio-quiet quasars/AGN compared to radio-loud ones emerges naturally. Once we acquire quantitative statements about the fraction of retrograde versus prograde configurations forming in mergers, the energetics required to turn FRII objects into FRIs, and the expected fraction of spin flips, we will be able to constrain the gap paradigm in relation to the observed 15–20% value of the radio-loud/radio-quiet dichotomy.

## 8. On Gamma-Ray Loud NLS1s and Jets in Spiral Galaxies

Until recently, spiral galaxies have generally entered into the AGN classification as radio-quiet objects which has prompted spin paradigm models to assume they have low- spinning black holes in their centers [26, 67, 68]. These ideas are grounded in the possibility of chaotic accretion whose purpose is to produce low-spinning black holes on average in such systems [91], but evidence is growing that the fraction of high-spinning prograde black holes is large in Seyferts [92, 93]. In addition, powerful jets have recently been discovered in spirals [66, 94–96]. Within the context of the spin paradigm, spiral AGN with powerful jets must contain high-spinning black holes. However, the Eddington ratio luminosities in these objects are difficult to reconcile with this notion because they appear to live in a regime of Eddington ratio luminosity that is intermediate between FSRQs and PG quasars [26, 96]. More precisely, FSRQs are observed to have lower Eddington ratios compared to gamma-ray loud NLS1s. In spin paradigm models, FSRQs and gamma-ray loud NLS1s should occupy the same region of the Eddington ratio luminosity because they would naturally be scale-free equivalents. In other words, if we model these two classes of AGN as prescribed in the spin paradigm, that is, as high prograde spinning black holes surrounded by truncated inner accretion disks, their powerful jets differ only because the black hole masses are larger in FSRQs. Therefore, their Eddington ratios would be expected to be equal. However, what we find is that on average gamma-ray loud NLS1s have larger Eddington ratios. This difference requires an explanation. In the gap paradigm, on the other hand, spiral AGN with jets are lower-spinning prograde systems so their disk efficiencies are naturally sandwiched between FSRQs (which are modeled as retrograde systems with smallest disk efficiencies) and radio-quiet AGN/quasars (which are modeled as the highest-spinning prograde systems, and thus with the largest disk efficiency). The notion that the overwhelming fraction of observed radio-quiet AGN as high-spinning prograde systems being due to a selection effect [93] stands in contrast to these intermediate-Eddington luminosities in gamma NLS1s.

## 9. On GRMHD

Numerical simulations of magnetized black hole accretion involve the general relativistic version of the Maxwell equations, which amounts to an advection and diffusive terms competing in the time evolution equation for the magnetic field, but this physics is not fully implemented in general relativistic simulations [97, 98]. Due to complications with the numerics, the diffusive term is absent. Hence, there is no physical diffusion in GRMHD (see Bucciantini and Del Zanna [99] for recent issues and progress on resistive GRMHD). This fact has received far less emphasis than it should. What is the impact of this absence of physics? Nonrelativistic simulations of protostellar disks including resistive MHD [100, 101] show that the so-called "non-ideal" terms—that is, the terms that capture diffusive effects—are as important to the outcome as the ideal ones. Given that general relativity involves more intense dynamical interaction between the gas and magnetic field, especially close to the black hole in the ergosphere [102, 103], diffusive terms should be at least as important there as in the Newtonian regime, and most likely more so by some order(s) of magnitude. Unfortunately, this issue is no longer mentioned in the GRMHD literature. In addition, recent work in plasma theory, suggests that even if GRMHD simulations did include the standard diffusive terms, they would still produce results that are off by orders of magnitude due to the so-called "stochastic flux- freezing" [104].

The other issue confronting numerical simulations of magnetized black hole accretion involves the nature and geometry of the magnetic field [105]. In the absence of some naturally expected large-scale magnetic field configuration in the nuclei of galaxies, the geometry must be put in by hand according to some ad hoc prescription. In fact, due to computational limits, the geometry in GRMHD simulations tends to be specified and restricted to the inner regions of the accretion disks, usually in the form of internal loops of magnetic field that produce a zero net flux, a highly unnatural state of affairs. This would not be much of an issue were it not for the fact that different initial magnetic field configurations produce sufficiently different black hole- threading magnetic fields and therefore different jet powers [106, 107]. The combination of these two issues—choice for magnetic field geometry and absence of physical diffusion— can yield unrealistic detailed dependencies such as on the black hole spin [82]. While certain dynamical aspects of actual astrophysical magnetized black hole accretion should be reproduced in GRMHD, the arguments suggest that detailed dependencies cannot yet be properly captured (see [105] Section 4), and it is within this context that analytic and semianalytic models remain essential.

**10. On Time Evolution**

The observed redshift distribution of radio galaxies and quasars provides a space for themost stringent constraints on theory [108, 109]. The gap paradigm postulates the presence of retrograde accretion in the most massive black hole mergers but not in spiral galaxies and black hole X-ray binaries, where the difficulty in forming a retrograde system constrains them to a prograde accretion regime. While the inverse relation between accretion and jet efficiency in the gap paradigm model has been the primary focus so far, the time evolution and consequent unification are arguably its most attractive features. This unification connects radio-loud objects (or sources with jets) to one another and to radio-quiet ones (or sources without jets) and in doing so produces an explanation for their redshift distribution. Despite claims in the recent literature [26, 67, 68], the spin paradigm continues to be at odds with observations in this respect, as discussed below.

The postmerger retrograde systems, that is, the starting assumption of the gap paradigm involve cold, radiatively efficient, and Shakura and Sunyaev type accretion with powerful jets, with cold accretion being the result of post-merger funneling of cold gas into the nuclear region [110]. The subsequent time evolution does not require additional assumptions. This point requires emphasis. Accretion simply spins the black hole down and then up again in the prograde regime, so the gap region decreases in size with time. On the foundation of that simple picture, the model proposes an explanation for the evolution of radio-loud quasars into radio-quiet quasars and/or radio galaxies. Accordingly, the fraction of FRII objects is larger at higher redshifts and gives way to a dominance of FRI objects at lower redshifts. An explanation is also found in this picture for the absence of merger signatures associated with FRI radio galaxies [111] in that they are the late-state evolution of originally retrograde black hole accretion systems that in turn were triggered by galaxy mergers. It is worth noting that on purely accretion timescales, the Eddington limited accretion will spin down to zero spin a maximally retrograde black hole in just under 10 million years. This timescale is beautifully and otherwise coincidentally compatible with estimates of FRII lifetimes [112]. Also, this framework provides a means for understanding that scaling relations should differ between radio-loud AGN and radio-quiet AGN, the former being determined by a combination of different types of jets (FRII/FRI) and accretion (radiatively efficient to radiatively inefficient), the latter being governed by disk winds in radiatively efficient accretion.

Spin paradigm models, on the other hand, struggle at the outset in modeling the powerful FRII quasars because they are forced to high spin and prograde accretion which maximizes both the jet and accretion efficiency. The already mentioned tension that such a combination of requirements produces is further compounded by its difficulty to explain the inverse relationship between accretion outflow power in radio-loud quasars versus radio-quiet quasars. In other words, accretion outflow power should be greater in sources that have high prograde spin compared to those with low spin, yet the sources with the most powerful jets invariably show distributions spanning a range of weaker disk outflows [113, 114]. Martinez-Sansigre & Rawlings [67] argue that it is the low-number statistics that creates this distribution and that once the number of sources increases, the wind power in radio-loud quasars will be shown to dominate over the radio- quiet population. But what would the observations have to be to support an inverse relationship between accretion wind power and radio-loud AGN? In other words, what would the observations have to be to support the gap paradigm? They would be exactly what they are. In addition, FRII quasars have on average larger black hole masses compared to radio-quiet quasars which means that observations of faster winds in radio-quiet quasars is even more statistically significant. Hence, the conclusion of Martinez-Sansigre & Rawlings is a hope, one that spin paradigm and hybrid spin paradigm models share explicitly or implicitly, but one that unfortunately is the hallmark of a nonfalsifiable framework.

The observations of low-luminosity FRII radio galaxies at intermediate redshifts between FRII quasars and low-luminosity FRI radio galaxies is a natural outcome of time evolution in the gap paradigm (as described in detail in [37]). The redshift distribution of low-luminosity FRII radio galaxies in the spin paradigm, on the other hand, finds no natural explanation in terms of time evolution, and certainly not in a scale-invariant sense. And, as pointed out above, the jet efficiency requirements [73, 75, 76] in these objects are currently unresolved within spin paradigm and hybrid spin paradigm models regardless of explanations that deal with their redshift distributions. The gap paradigm, again, produces much less stringent requirements on the accretion rate [79].

Because retrograde systems may be unstable [115], and naturally tend to prograde systems, a possible explanation to the observation that radio-loud AGN host the most massive black holes ([116] and references therein) emerges in the gap paradigm. In fact, the larger the black holes, the more stable their retrograde accretion phase, while smaller black holes would tend to be less stable. The picture in the gap paradigm would go as follows: any high-spinning retrograde accretion system with a less massive black hole may flip to a prograde configuration, thereby becoming a radio-quiet quasar or AGN. The transition from the retrograde to the prograde regime is also an attractive place to explore explanations for X-shaped radio galaxies and the missing objects of the so- called "blazar envelope" [117].

**11. Conclusions**

Despite a common foundation grounded in accretion and black hole spin, the gap paradigm and spin and hybrid spin paradigms (SHSPs) differ in substantial ways. They are listed below. The emphasis of this paper has been on the observations and the extent to which theoretical differences can currently be constrained.

(1) In the gap paradigm, powerful FRII quasars involve retrograde accretion onto rotating black holes; while in SHSPs, FRII quasars involve prograde accretion onto rotating black holes.

(2) In the gap paradigm, FRII quasars evolve into FRII radio galaxies or FRI quasars and eventually either into FRI radio galaxies or radio-quiet quasars; while in SHSPs, FRII quasar evolution has no clear or model-constrained scale-invariant progression.

(3) In the gap paradigm, FRI quasars live in a range of lower prograde black hole spin values; whereas in SHSPs, they are high-spinning prograde systems.

(4) Whereas in the gap paradigm radio-quiet quasars are high-spin prograde systems surrounded by cold, thin-disk accretion, in SHSPs, they are low-spin systems surrounded by cold, thin-disk accretion.

(5) Whereas in the gap paradigm FRII objects will have a greater density at higher redshifts compared to FRI objects, in SHSPs, the natural scale invariant approach provides no redshift-dependent difference between FRI and FRIIs.

(6) Whereas in the gap paradigm spiral galaxies generate smaller-scale equivalents of FRI quasars, in SHSPs, spiral galaxies generally do not produce jets because they tend to include the assumption of secular processes and chaotic accretion, but when they do have jets, the spin must be prograde and high.

(7) Whereas in the gap paradigm the distribution of the black hole spin in the local universe should be prograde because such is the natural outcome of prolonged accretion, in SHSPs, the distribution of the black hole spin in the local universe should be around zero.

(8) Whereas in the gap paradigm X-shaped radio galaxies may find an explanation at the transition between retrograde and prograde systems, in SHSPs, there are no natural explanations for their presence along the boundary of the Owen- Ledlow diagram.

(9) Whereas in the gap paradigm FSRQs should have lower-Eddington ratios compared to gamma NLS1s, in SHSPs, their Eddington ratios should be equivalent.

Acknowledgment

The author thanks Rob Fender for setting him straight on a number of issues.